\documentclass[a4page,graphicx,12pt]{elsarticle}
\usepackage{graphics}
\usepackage{epsfig}
\usepackage{graphicx}
\usepackage{color,soul}

\begin{document}
\begin{frontmatter}
\title{\bf Magnetism on Rough Surfaces of Fe, Co and Ni : An Augmented Space Approach}
\author[label1]{Priyadarshini Parida}
\author[label1]{Biplab Ganguli\corref{mycorrespondingauthor}}
\cortext[mycorrespondingauthor]{Corresponding author}
\ead{biplabg@nitrkl.ac.in}
\author[label2,label3,label4] {Abhijit Mookerjee}
\ead{abhijit.mookerjee61@gmail.com}
\address[label1]{Department of Physics, National Institute of Technology Rourkela - 769008, India.}
\address[label2] {Department of Physics and Materials Science, S.N. Bose National Centre for Basic Sciences, JD-III Salt Lake, Kolkata - 700098, India.}
\address[label3]{Distinguished Visiting Professor, Presidency University, College Street, Kolkata, India}
\address[label4]{Visiting Professor, Lady Brabourne College, Suhrawardy Avenue, Kolkata, India}

\begin{abstract}
The augmented space formalism coupled with the recursion method and a tight-binding linear Muffin-tin orbitals basis has been applied to study the effects of roughness on the properties of (001) surfaces of body-centered cubic Fe and face-centered cubic Co and Ni. The formalism is also proposed for the study of smooth surface. Comparisons have been made for three types of surfaces: a smooth surface, the surface with a rough top layer, and a more realistic model with several rough top layers converging into a crystalline bulk. Comparisons have been made between the magnetic moments, work function and electronic density of states in the three models described above. 
\end{abstract}

\end{frontmatter}

\section{Introduction}

When a film is deposited layer by layer during an epitaxial growth, there is always a possibility of vacancies being created in the topmost layers. There are extensive studies on surface magnetism of ideal surfaces for over three decades. But a realistic surface is actually rough which forms during the process of epitaxial growth. Therefore, it is important to model a rough surface in any theoretical studies in realistic case. The combined effects of randomness in the incoming atomic flux and the smoothening due to surface diffusion in finally creating rough surfaces and overlayers have been studied earlier \cite{Mehta, Huda}. The top layers are expected to have maximum roughness as the atoms in these layers have looser bonding, lower coordination number and reduced symmetry as compared to bulk atoms. Rough surfaces can be modelled as a random binary alloys. We can envisage the missing atoms at a fraction of sites in the surface layers as random occupation of these sites by ``empty spheres". These spheres are hardly ``empty". They lack ion-cores, but carry non-negligible electronic charge. In the LMTO technique such empty spheres are introduced to take care of large interstitial space in non-compact structures. Consequently, when we solve the Kohn-Sham equations within the local spin-density functional approximation (LSDA), the 
empty spheres are an integral part of the self-consistent treatment. We model our rough surface as
a substitutional binary disordered alloy of the surface atoms carrying charge and ion-cores and empty spheres with charge alone. Due to these roughness, there is significant change in physical properties of a surface as compared to the bulk. 

We have taken two layers of empty spheres on top of the surface layer. This was to take care of charge leakage into the vacuum. Because of reduced co-ordination number at the surface, surface atoms are weakly bound compared to atoms in the bulk and so the interatomic separation is expected to be slightly different than in the bulk. Therefore lattice relaxation is necessary. In this communication we consider lattice relaxation of surface layers. A different relaxation procedure was adopted for Fe(001) where interlayer gap is relaxed using FPLAPW \cite{freeman1}.

To understand surface properties, it is very important to know the surface morphology. In transition metals, itinerant Fermi electrons play an important role in determining surface properties. In such systems, magnetic ordering arises due to the interaction among these Fermi electrons. Surface electronic and magnetic properties of such systems depend on the detailed electronic structure, the atomic arrangement and the composition.

Our primary aim here to model realistic rough surface using Augmented space formalism (ASF) but we have also shown that ASF can be extended to include ideal smooth surface. Therefore we have also carried out studies on ideal surface for comparison purpose with existing studies.

\section{Computational Methods}

Disorder has been tackled earlier with both the crude virtual crystal approximation and for a more
sophisticated and self-consistent mean-field theory approach - the coherent potential approximation
(CPA). The drawbacks of the CPA were immediately recognized and the search was on for going
beyond mean-fields. The ASF \cite{mookerjee, mookerjee1} emerged as one of the very few methods that accurately described not only the averaged background disorder, but also a large deal of 
configuration fluctuations. The ASF has been described in detail in a number of papers \cite{mookerjee, mookerjee1,rudra} and book \cite{TF}. We refer the reader to them and mention here only those aspects of it which are relevant to our problem. In a nutshell, the ASR replaces the random parameters of the Hamiltonian by operators whose eigenvalues were the values taken by the parameters with 
probability densities which are the spectral densities of these operators. The augmented space
theorem\cite{mookerjee1} then showed that the configuration average a special matrix element in
configuration space.
                                          
\[ n(R) \Rightarrow {\cal N}(R) = \left(\begin{array}{ll}    y & \sqrt{xy}\\
                                \sqrt{xy} & x 
                                \end{array}\right)   \] 
                                                         
The right-hand side is the representation of an operator $\cal{N}$ in the configuration space $\Phi$. The augmented Hamiltonian is

\begin{eqnarray}
\tilde{H} &= &\sum_{RL} \left( C^B P_{RL}\otimes I + (C^A-C^B) P_{RL}\otimes N_R \right) + \ldots\nonumber \\
& + & \sum_{RL,R'L'} \left( \Delta^B P_{RL}\otimes I + (\Delta^A-\Delta^B) P_{RL}\otimes N_R \right)\otimes \ldots\nonumber\\
&  \otimes & \left(S_{RL,R'L'} T_{RL,R'L'}\otimes I\right)\otimes \left( \Delta^B P_{R'L'}\otimes I + (\Delta^A-\Delta^B) P_{R'L'}\otimes N_{R'} \right)
\end{eqnarray}

Here $P$ and $T$ are projection and transfer operators. $R$ are the lattice sites and $L=(lm)$ are the orbital indices. For transition metal $l<2$. $C^A$, $C^B$, $\Delta^A$ and $\Delta^B$ are the TB-LMTO potential parameters of the constituents atoms $A$ and $B$ of the alloy. $S$ is the structure constant.

The augmented space theorem \cite{mookerjee} gives the configuration averaged Green function as :

\begin{equation}
\ll G_{RL,R'L'}(z) \gg = \langle RL\otimes\emptyset| (z\tilde{I}-\tilde{H})^{-1}|R'L'\otimes\emptyset\rangle 
\end{equation} 

The Green function is then obtained as a continued function expansion through the Recur sionMethod of Haydock et.al. \cite{haydock} and is generalized for operators on the real space augmented with the space of configuration fluctuations. It involves the familiar three term recursion :

\begin{eqnarray}
|1\rangle = |R\otimes \emptyset\rangle \qquad |2\rangle = \tilde{H} |1\rangle - \alpha_1|1\rangle\nonumber\\
|n+1\rangle = \tilde{H} |n\rangle - \alpha_{n} |n\rangle - \beta^2_{n-1}|n-1\rangle \nonumber \\
\alpha_n = \langle n|\tilde{H}|n\rangle/\langle n|n\rangle \qquad \beta^2_n = \langle n|n\rangle/\langle n-1|n-1\rangle
\end{eqnarray}

Leading to :

\begin{equation}
 \ll G_{RR}(z) \gg \ =\ \frac{\beta^2_1}{\displaystyle z-\alpha_1-\frac{\beta_2^2}{\displaystyle
 z - \alpha_2 - \frac{\beta_3^2}{\displaystyle z-\alpha_3-\frac{\beta_4^2}{\displaystyle \ddots T(z)}}}}
 \end{equation}

$\alpha$ and $\beta$ are potential parameters and $T(z)$ is the terminator.

The potential parameters are generated from TB-LMTO within local spin density approximation (LSDA) using Barth and Hedin exchange correlation potential. We have used seven shell augmented space calculation and nine steps of recursion. 

\section{Results for Rough Surfaces}

We have chosen twelve atomic layers. The top most layer is relaxed by the 5\% increment of lattice constant in the case of bcc Fe(001) whereas it is respectively 16\% and 9\% in the case of fcc Co(001) and fcc Ni(001). We have considered two types of surfaces, one by roughening the top most layer and the other by roughening the top four layers with different degree of roughness. This figure of percentage of increment are obtained by minimizing the total energy.

\begin{figure}[p]
\centering
\includegraphics[scale=0.6]{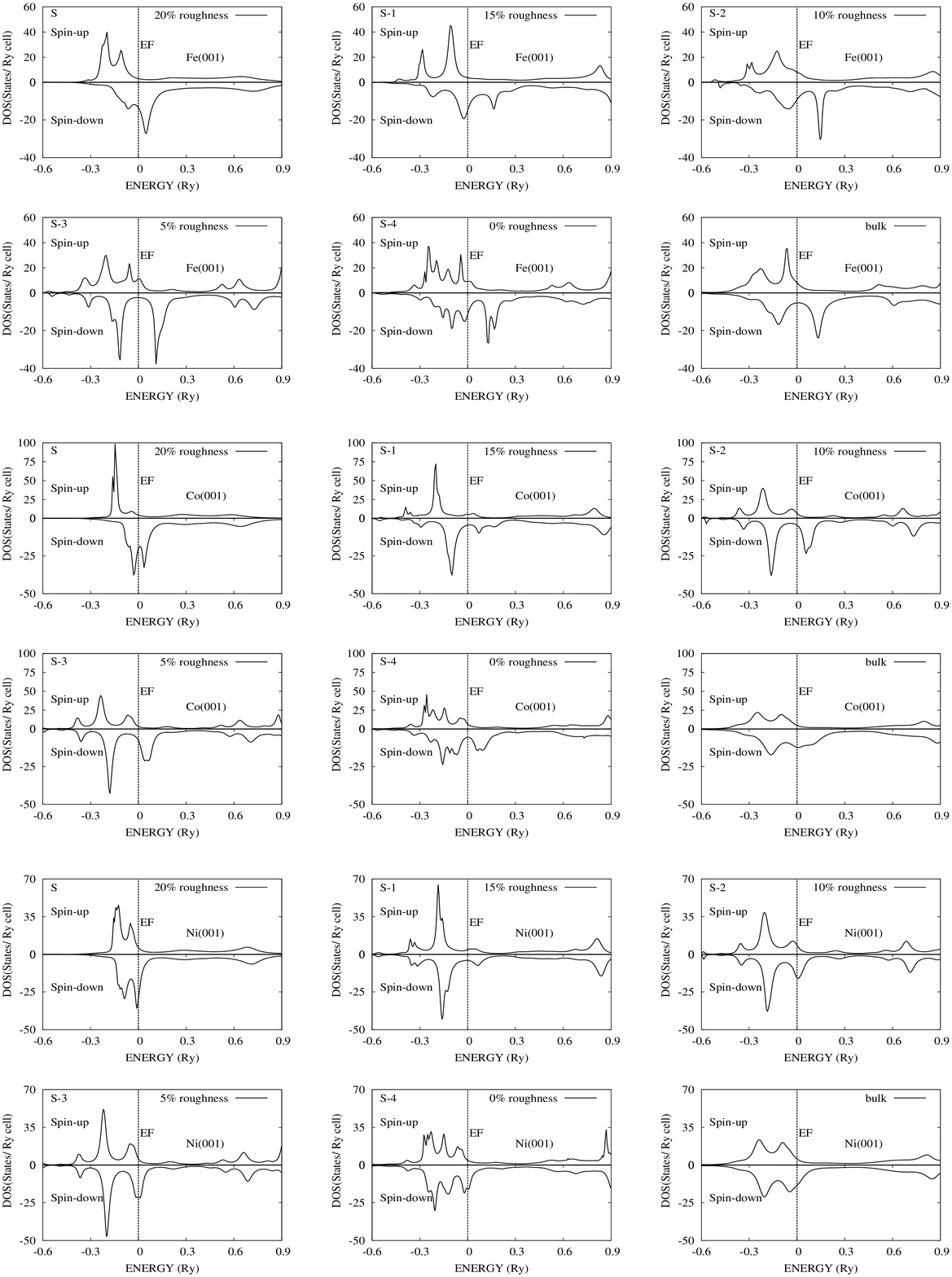}
\caption{Spin resolved density of states starting from the surface layer down to the bulk along the (001) direction. The roughness also decreases from 20\% on the surface to none in the bulk. Fermi level is reset at zero.}
\label{fig1}
\end{figure}

 Roughening is considered from 20\% to 5\% from top layer to the fourth lower layer with a difference of 5\%. Figure \ref{fig1} shows as we go down from the layer S to the bulk, width of  density of states (DOS) increases. In the case of surfaces the width of the DOS of the top most surface layer (S) is narrower as compared to the bulk which is expected. The width increases substantially at S-1 layer and then reaches to the bulk values very slowly. Though width of DOS from S-2 layer onwards changes slowly, but there are significant changes in structure of DOS. This is due to variation in roughness. DOS bulk value is reached at the S-9$^{th}$ layer down the top most layer in the case of Fe(001) whereas it is reached at the S-8$^{th}$ layer in case of Co(001) and Ni(001). 

Figure \ref{fig1} also shows interesting variation in spin resolved DOS. We note that down spin electrons contribute significantly to DOS at the Fermi level at the layers S \& S-1 in the case of Fe(001). This trend is changed at the S-2 level where both up and down spins have significant contributions indicating that this layer is less magnetic. It is other way round at S-3 level where up has significant contribution and down is negligible. In the case of Co(001) contributions of spin-down to the DOS at the Fermi level is significant compared to spin-up at all the layers. This is also true in the case of Ni(001) except at S-1 layer where both have significant contributions. This means S-1 layer of Ni is almost non-magnetic at Fermi level. The effect of roughness is visible in this figure. For example in the case of Fe the width of DOS at S-4 layer where roughness is zero is more compared to other layer where roughness increases. Apart from the difference in width, the roughness smoothen the curve.

\begin{table}
\caption{Work Functions.\label{tab1}}
\centering
\begin{tabular}{@{}||c||c|c|c||c|c|c||} \hline

&\multicolumn{3}{c||}{Four layered rough surface} & \multicolumn{3}{c||}{Smooth surface}\\ \cline{2-7}

\multicolumn{1}{||c||}{\raisebox{1.2ex}[0pt]{Properties}} & Fe(001) & Co(001) & Ni(001)& Fe(001) & Co(001) & Ni(001) \\ \hline
Work Function (eV) & 4.15 & 5.42 & 4.79& 4.15 & 5.33 & 4.79\\ \hline
\end{tabular}
\end{table}

Table \ref{tab1} shows the work function.  There is negligible effect of roughness on work function for the case of Fe and Ni but there is slight change in the case of Co.

\begin{table}
\caption{Layered based and bulk (B) orbital resolved magnetic moment in $\mu_B$/atom.\label{tab2}}
\centering
\begin{tabular}{@{}||c||c||c|c|c|c||c|c|c|c||c|c|c|c||}
\hline

& Roughness & \multicolumn{4}{c||}{Fe(001)}  & \multicolumn{4}{c||}{Co(001)} &\multicolumn{4}{c||}{Ni(001)} \\ \cline{3-14}
\multicolumn{1}{||c||}{\raisebox{1.2ex}[0pt]{Layers}} &(\%) & s & p & d & Total & s & p & d & Total & s & p & d & Total \\ \hline
S & 20 & -0.06 & 0.03 & 2.82 & 2.79  & -0.01 & 0.03 & 2.30 & 2.32 & 0.0 & 0.0 & 0.63 & 0.63 \\\hline
S-1 & 15 & -0.05  & -0.03 & 2.15 & 2.07 & -0.07 & -0.01 & 1.20 & 1.12 & -0.02 & 0.0 & 0.43 & 0.41 \\ \hline
S-2 & 10 & -0.02 & -0.03 & 1.89 & 1.84 & -0.02 & -0.04 & 1.82 & 1.76 & -0.01 & -0.01 & 0.64 & 0.62 \\\hline
S-3 & 5 & -0.02 & -0.05 & 2.13 & 2.06 & -0.01 & -0.05 & 1.97 & 1.91 & -0.01 & -0.02 & 0.74 & 0.71 \\ \hline
S-4 & 0 & -0.01 & 0.0 & 2.02 & 2.01 & 0.04 & -0.06 & 1.72 & 1.70 & 0.01 & -0.02 & 0.36 & 0.35 \\\hline 
B & 0 & -0.02 & -0.06 & 2.25 & 2.17 & -0.02 & -0.06 & 1.62 & 1.54 & -0.01 & -0.02 & 0.55 & 0.52\\ \hline
\end{tabular} 
\end{table}

\begin{figure}[t]
\centering
\includegraphics[scale=0.6]{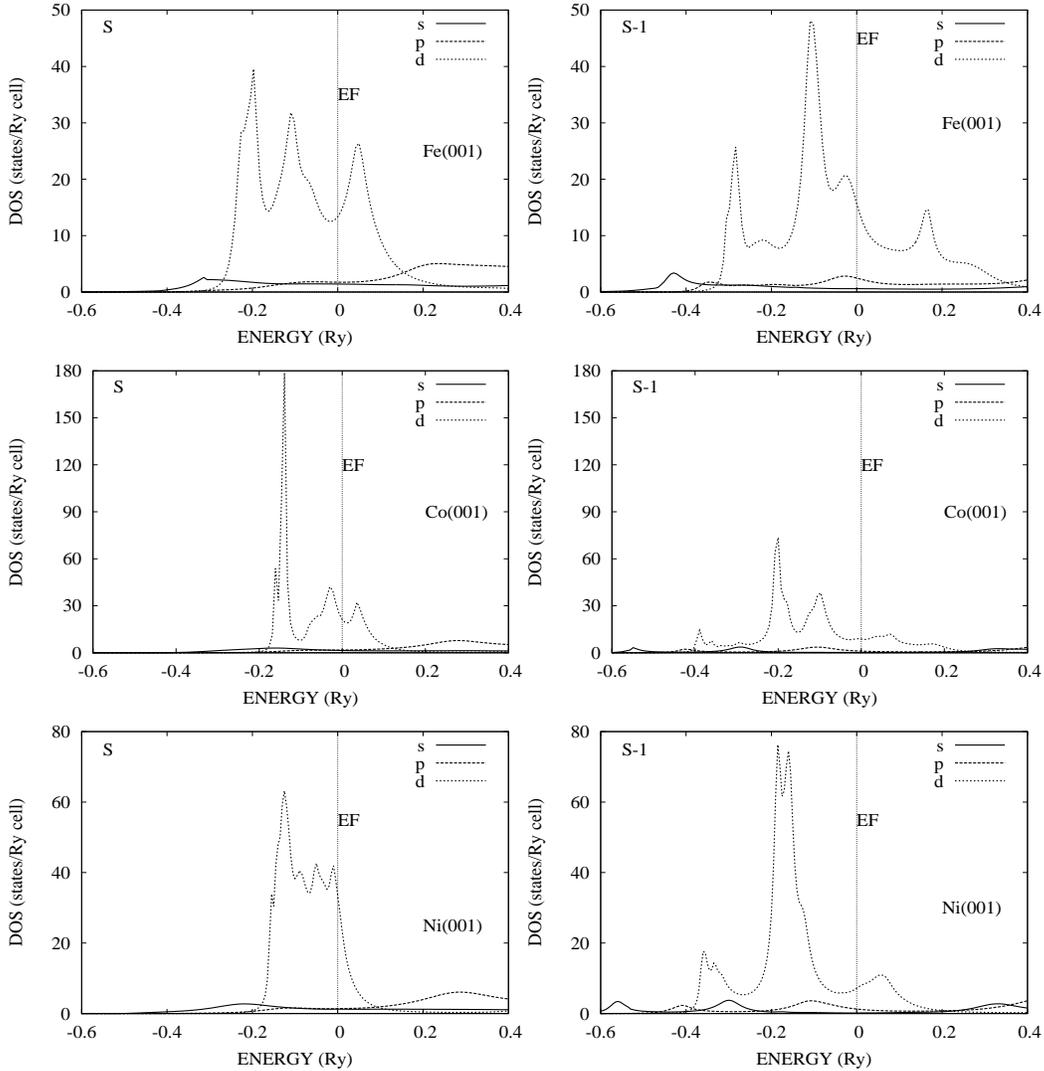}
\caption{Orbital resolved total DOS for surface (S) and sub-surface (S-1) layers. Fermi level is reset at zero.}
\label{fig2}
\end{figure}

Table \ref{tab2} shows that the topmost layer (S) of Fe(001) \& Co(001) has the maximum magnetic moment. Whereas layer S-3 of Ni(001) has maximum magnetic moment. This is indeed true because splitting of spin up and down DOS is maximum at these respective layers as shown in the Figure \ref{fig1}. The bulk magnetic moment is attained at the S-9$^{th}$ layer in the case of bcc Fe(001) whereas it is attained at the S-8$^{th}$ layer in the case fcc Co(001) and fcc Ni(001). The average magnetic moment of the top four layers having different amount of roughness is more than its bulk value for all the three systems. This is expected as surface magnetic moment is enhanced compared to the bulk. The table \ref{tab2} shows d-orbital alone contributes most to the the magnetic moment as expected. Similarly, Figure \ref{fig2} shows d-orbital DOS is maximum whereas others are almost negligible. Among the three systems, the most significant change in d-band DOS of S layer is in Co(001). Therefore its magnetic moment in the S$^{th}$ layer is almost double that of S-1 layer.

\begin{figure}[p]
\centering
\includegraphics[scale=0.75]{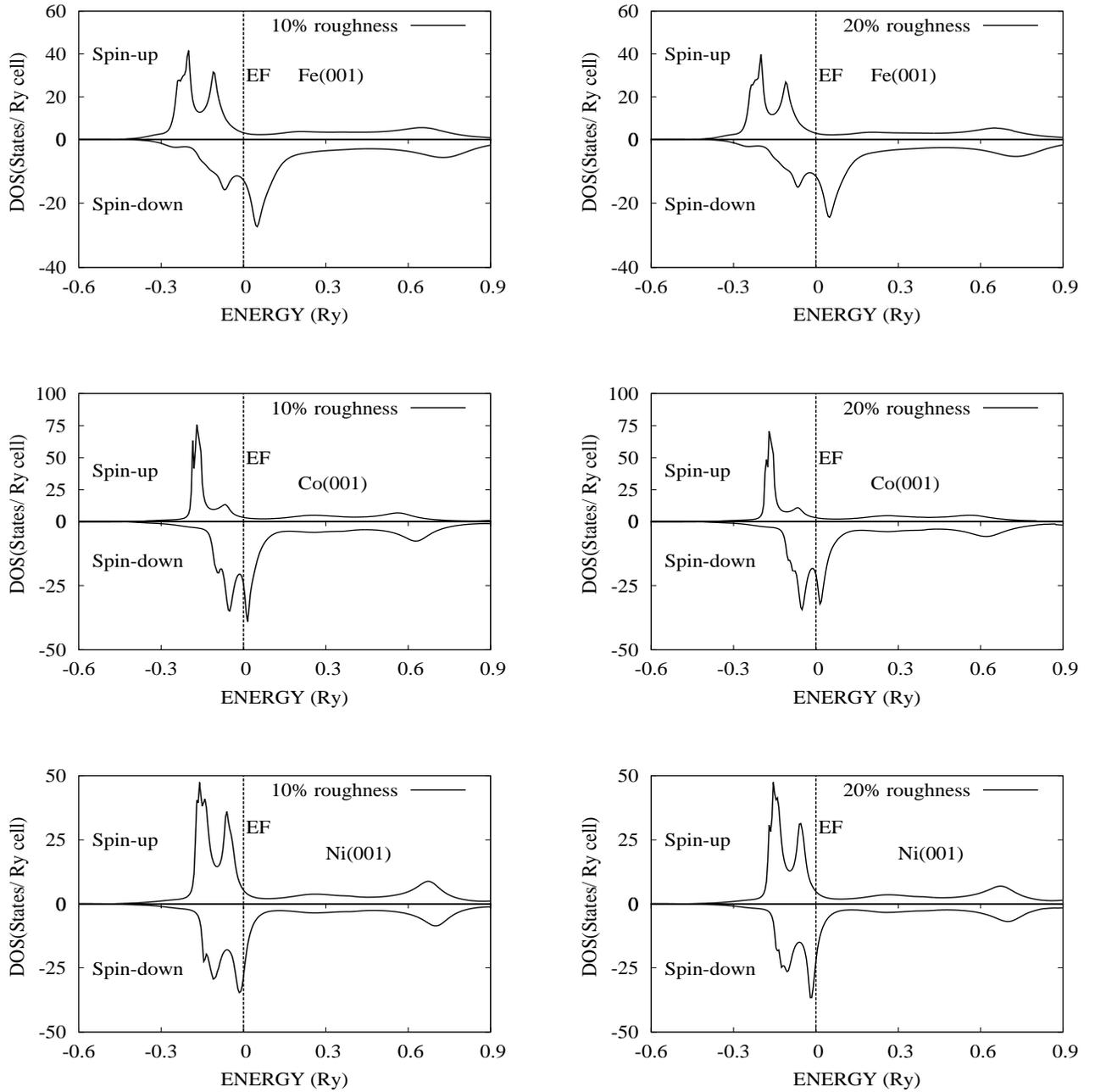}
\caption{DOS of top layer (S) of Fe(001, Co(001) and Ni(001) in a model in which only this layer has been roughened only.  Fermi level is reset at zero.}
\label{fig3}
\end{figure}

Majority of the work on surface properties is the study of rough topmost layer due to poisoning of substrate atoms \cite{biplab,biplab1}, not due to voids. Therefore, we had roughened the top most layer with different degrees of roughness, (with 10\% \& 20\% of empty spheres). Figure \ref{fig3} shows that the peak near the Fermi level changes significantly when roughness is varied in all the three systems. Hence the electrons near the Fermi level are more affected by the change in roughness. It is found that the average magnetic moment decreases for all the systems when roughness is increased. When the roughness is more than 50\%  the surface magnetic moment increases and it approaches to the atomic magnetic moment for these systems. It is also observed that work function of the top layer changes slightly when roughness is varied. This value remains almost equivalent to smooth surface values. In this case also we observe the magnetic moment of the top most layer more than all other layers for Fe(001) and Co(001) where as layer S-3 has more magnetic moment than other layers for Ni(001). 

Comparison between the DOS of the two cases, i.e., a realistic case when first four layers are roughened (Figure 1) and the other when only the top layer is roughened with 20\% (Figure 3) indicates that there is minor difference in DOS in the case of Fe(001) and Ni(001). But there is significant difference in the case of Co(001). Similarly comparison of Tables \ref{tab2} \&  \ref{tab3} show magnetic moments get enhanced significantly for the realistic case than for the other in the case of Co(001), where as there is only minor difference in the other two systems.

\begin{table}
\caption{Layer based and bulk orbital resolved magnetic moment when only top layer is roughened with 20\% in $\mu_B$/atom.}
\label{tab3}
\centering
\begin{tabular}{@{}||c||c|c|c|c||c|c|c|c||c|c|c|c||}
\hline
& \multicolumn{4}{c||}{Fe(001)}  & \multicolumn{4}{c||}{Co(001)} &\multicolumn{4}{c||}{Ni(001)} \\ \cline{2-13}
\multicolumn{1}{||c||}{\raisebox{1.2ex}[0pt]{Layers}} & s & p & d & Total & s & p & d & Total & s & p & d & Total \\ \hline
S & 0.01 & 0.03 & 2.81 & 2.85 & -0.01 & 0.03 & 1.93 & 1.95 & -0.01 & 0.0 & 0.65 & 0.64 \\ \hline
S-1 & 0.0 & 0.0 & 1.27 & 1.27 & 0.02 & -0.03 & 0.92 & 0.91 & 0.02 & -0.01 & 0.29 & 0.30 \\ \hline
S-2 & -0.03 & -0.05 & 1.82 & 1.74 & -0.07 & -0.05 & 1.75 & 1.63 & -0.02 & -0.05 & 0.47 & 0.40 \\\hline
S-3 & -0.03 & -0.04 & 2.33 & 2.26 & -0.02 & -0.05 & 1.74 & 1.67 & 0.01 & -0.03 & 0.76 & 0.74 \\\hline
S-4 & -0.03 & -0.05 & 2.64 & 2.56 & -0.01 & -0.06 & 1.65 & 1.58 & 0.0 & -0.02 & 0.55 & 0.53 \\\hline
B & -0.02 & -0.06 & 2.25 & 2.17 & -0.02 & -0.06 & 1.62 & 1.54 & -0.01 & -0.02 & 0.55 & 0.52 \\\hline
\end{tabular} 
\end{table}

\section{Transition to smooth surfaces}

One would have thought, given the way we modelled the roughness of surfaces, that if we simply let the concentration of the empty spheres go to zero and we would recover the smooth surfaces. However, that is not the case. As the concentration of the empty sphere decreases, these `impurities' become more and more isolated and form highly spikey impurity states. The coherent potential approximation for example fails in this composition range and do not adequately reproduce the impurity structures in the density of states. Originally it was also thought that the ASR too misses out these structures. However, careful analysis of the ``terminator" or the asymptotic behaviour of the continued fraction, indicated that this reproduces impurity peaks quite accurately. We have to incorporate not only the singularities at the band edges, but also those lying on the compact spectrum of $H$. Viswanath and M\"uller \cite{viswa,viswa1} has proposed a terminator:

\begin{equation}  
T(z) = \frac{2\pi(E_m)^{(p+2q+1)/2}}{B\left(\displaystyle{\frac{p+1}{2}},1+q\right)} \vert z-E_0\vert^p\left\{(z-E_1)(E_2-z)\right\}^q 
\end{equation}

The spectral bounds are at $E_1,E_2$ with square-root singularities, $E_m^2 = E_1E_2$ and there is a cusp singularity at $E_0$ if $p=1,q=1$ or infra-red divergence if $p=-1/2,q=0$. $E_0$ sits on the compact spectrum of $H$. Magnus \cite{magnus} has cited a closed form of the convergent continued fraction coefficients of the terminator :

\begin{eqnarray}
\beta^2_{2n} & = & E_m^2\ \frac{4n(n+q)}{(4n+2q+p-1)(4n+2q+p+1)} \nonumber\\
\beta^2_{2n+1} & = & E_m^2\ \frac{(2n+2p+1)(2n+2q+p+1)}{(4n+2q+p+1)(4n+2q+p+3)}\nonumber\\
\end{eqnarray}

\begin{figure}[h]
\centering
\includegraphics[scale=0.6]{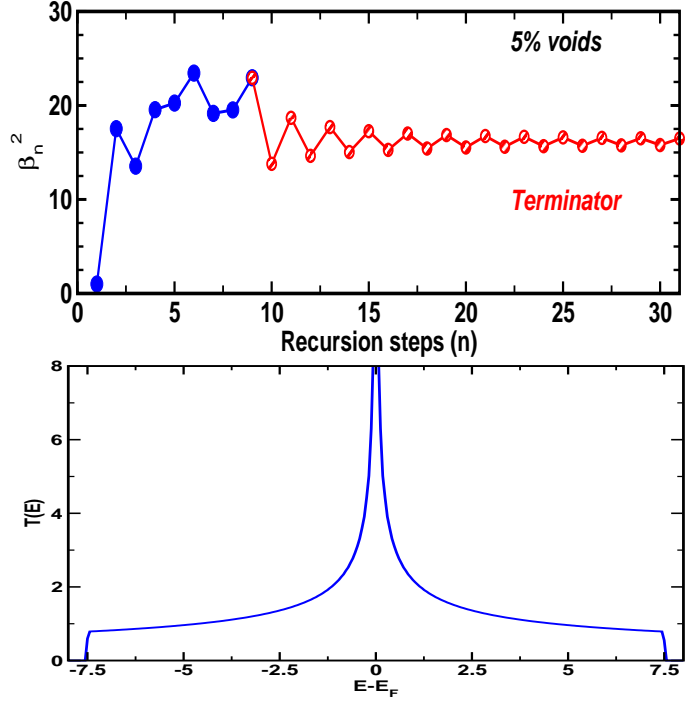}
\caption{(Top) Recursion coefficients $\beta_n^2$ from the recursive calculations
(blue) and terminator (red) smoothly enmeshed. (Bottom) Density of states with a peak at the origin\label{fig4}}
\end{figure}

The parameters of the terminator are estimated from the asymptotic part of the continued fraction coefficients calculated from our recursion. We shall use the Viswanath-M\"uller termination appropriate for infra-red divergences and seamlessly enmeshed with the calculated coefficients as shown in Figure \ref{fig4}.

\section{Discussions on almost Smooth Surfaces}

\begin{figure}[h]
\centering
\includegraphics[height=5cm,width=18cm]{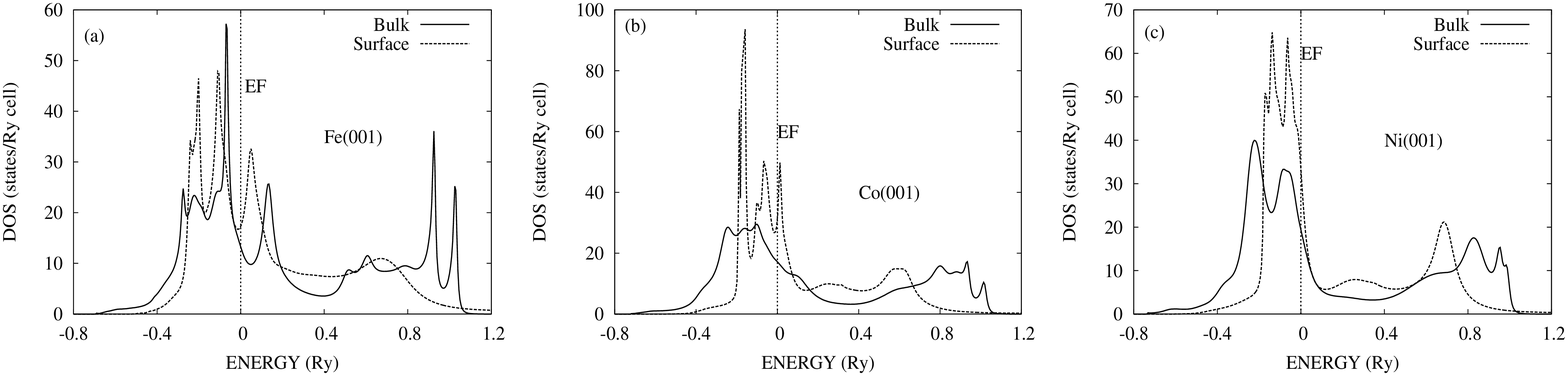}
\caption{Comparison of total density of states for bulk and surface (top most layer). Fermi level is reset at zero.\label{fig5}}
\end{figure}

\begin{figure}[h]
\centering
\includegraphics[width=18cm,height=6cm]{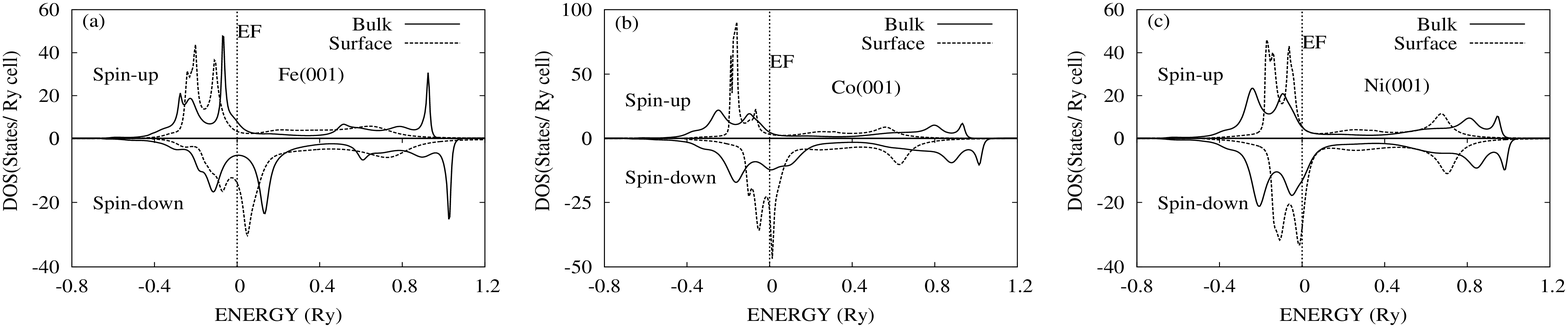}
\caption{Comparison of spin resolved density of states for bulk and surface (top most layer).  Fermi level is reset at zero.\label{fig6}}
\end{figure}

Figure \ref{fig5} shows energy bands of the surface states near Fermi level are narrower compared to bulk as expected. This is due to weakening of interaction by symmetry breaking and reduction in co-ordination number. Similar picture arise in the spin resolved DOS as shown in the Figure \ref{fig6}. Apart from narrowing of band there is change in number of spin-up surface states compared to the bulk states at the Fermi level. The amount of change is maximum in the case of Fe(001) and negligible in the case of Ni(001). But in all the three cases, there is significant change in the spin-down states. The splitting of spin-up and spin-down states near Fermi level is maximum for Fe(001) and least for Ni(001). This is expected because magnetic moment of Fe is maximum and it is least for Ni. The magnetic moment is directly related to the amount of splitting in spin up and spin down sates.
 
\begin{figure}[h]
\centering
\includegraphics[width=19cm,height=5.5cm]{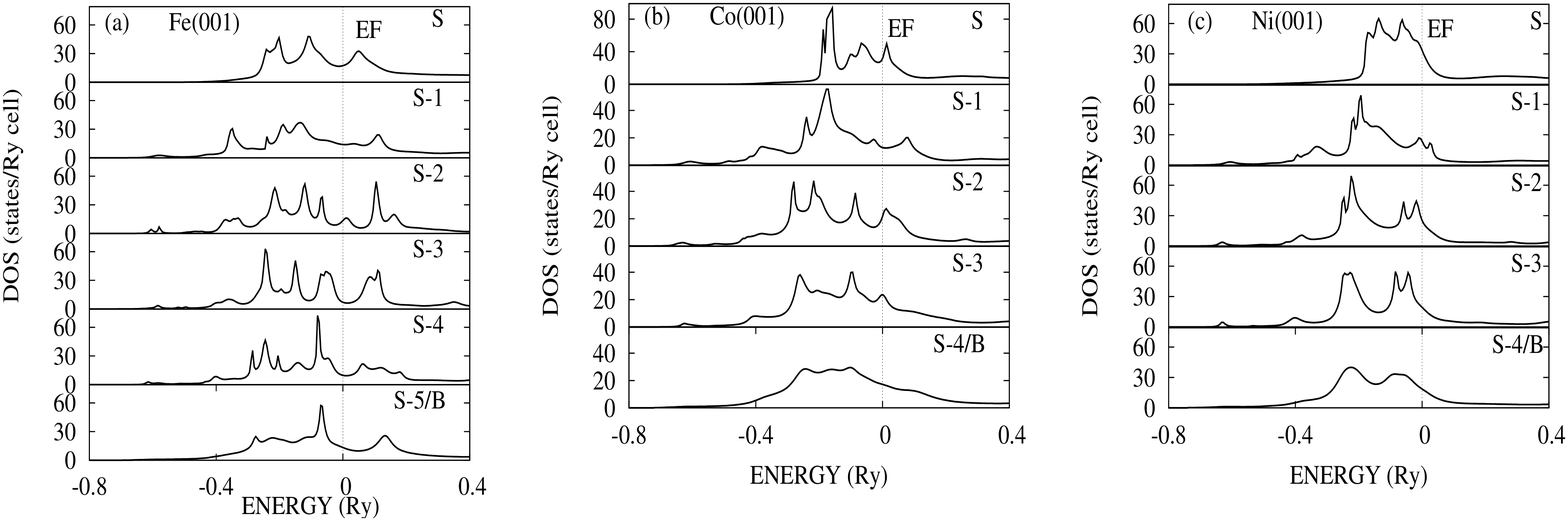} 
\caption{Comparison of layer based total DOS for surface (S), sub-surfaces (S-1, S-2, S-3, S-4, S-5) and bulk (B). Fermi level is reset at zero.\label{fig7}}
\end{figure}

The layer based total DOS, shown in the Figure \ref{fig7} indicates that DOS approaches the bulk value at S-5$^{th}$ layer in the case of Fe(001), whereas, it approaches to the bulk at S-4$^{th}$ layer in the other two, Co(001) and Ni(001), cases. On comparison of figures \ref{fig1} and \ref{fig7} we observe that the variation in structure in DOS between layers near the bulk is negligible in the case of smooth surface. The appearance of sharp peaks in DOS in figure \ref{fig1} compared to figure \ref{fig7} are due to disorderedness. For example, in the case of Co(001) 20\% produces very sharp spin-up peak compared to smooth surface.

\begin{figure}[h]
\centering
\includegraphics[scale=0.7]{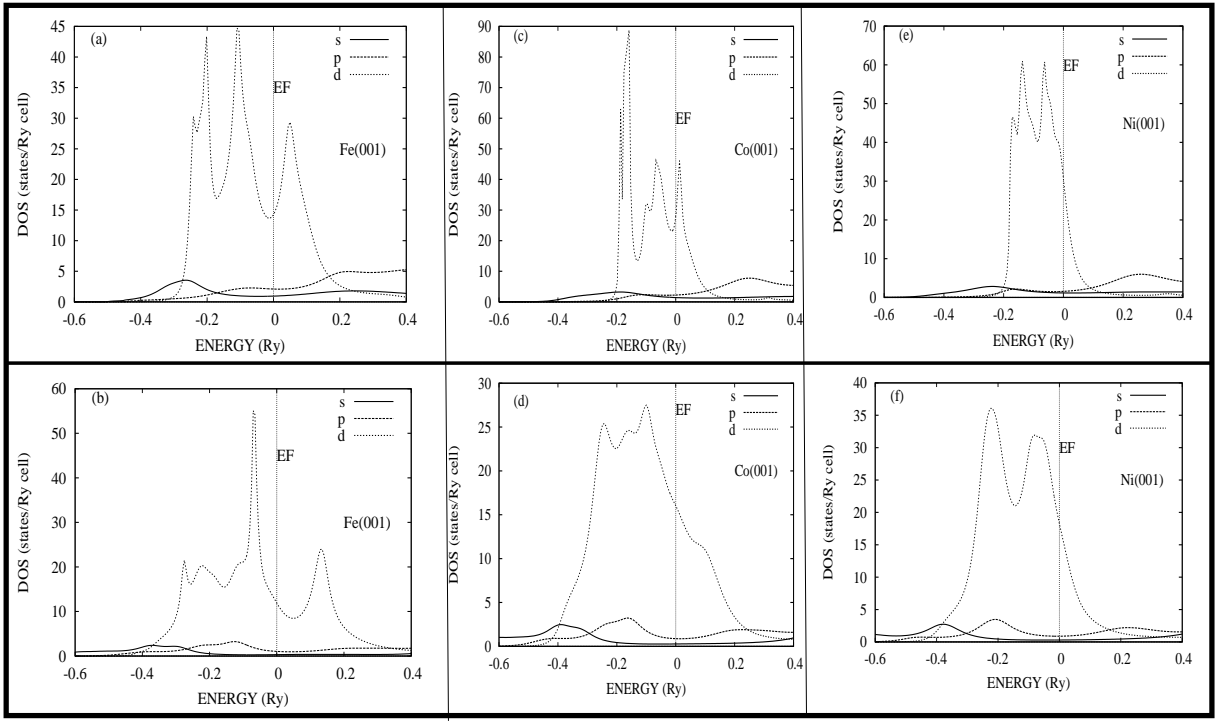}
\caption{Orbital resolved DOS: surface : (a), (c) \& (e) and bulk : (b), (d) \& (f). Fermi level is reset at zero.\label{fig8}}
\end{figure} 

In transition metals, d-electrons contributes mainly to the formation of bands. Therefore, any surface effects must be reflected into d-band. This is evident from the Figure\ref{fig8}, orbital resolved DOS. Apart from narrowing down of d-band, the band also gets splitted into few more prominent peaks at the surface .The narrowing of d-band is due to the dehybridization of s-, p-, and d-electrons \cite{freeman2}.  

\begin{table}
\caption{Orbital resolved magnetic moment for surface (S), sub-surfaces (S-1, S-2, S-3, S-4) and bulk (B) in $\mu_B$/atom.}\label{tab4}
\centering
\begin{tabular}{@{}||c||c|c|c|c||c|c|c|c||c|c|c|c||}
\hline

& \multicolumn{4}{c||}{Fe(001)}  & \multicolumn{4}{c||}{Co(001)} &\multicolumn{4}{c||}{Ni(001)} \\ \cline{2-13}
\multicolumn{1}{||c||}{\raisebox{1.2ex}[0pt]{Layers}} & s & p & d & Total & s & p & d & Total & s & p & d & Total \\ \hline

S & 0.01 & 0.02 & 2.80 & 2.83 & -0.01 & 0.02 & 1.82 & 1.83& -0.01 & 0.0 & 0.69 & 0.68\\\hline
S-1 & -0.02  & -0.03 & 2.13 & 2.08 & -0.01 & -0.03 & 1.49 & 1.45 & -0.01 & -0.01 & 0.60 & 0.58\\\hline
S-2 & -0.03 & -0.06 & 2.06 & 1.97& -0.01 & -0.04 & 1.79 & 1.74& 0.0 & -0.02 & 0.62 & 0.60\\ \hline
S-3 & -0.02 & -0.04 & 2.70 & 2.64 & -0.01 & -0.06 & 1.66 & 1.59& 0.0 & -0.02 & 0.56 & 0.54 \\ \hline
S-4 & -0.01 & -0.06 & 2.43 & 2.36&&&&&&&&\\\cline{1-5}
S-5/B & -0.02 & -0.06 & 2.25 & 2.17 & {\raisebox{1.2ex}[0pt]{-0.02}} & {\raisebox{1.2ex}[0pt]{-0.06}} & {\raisebox{1.2ex}[0pt]{1.62}} & {\raisebox{1.2ex}[0pt]{1.54}}& {\raisebox{1.2ex}[0pt]{-0.01}} & {\raisebox{1.2ex}[0pt]{-0.02}} & {\raisebox{1.2ex}[0pt]{0.55}} & {\raisebox{1.2ex}[0pt]{0.52}}\\\hline
\end{tabular} 
\end{table}

\begin{figure}[h]
\centering
\includegraphics[scale=0.4,angle=-90]{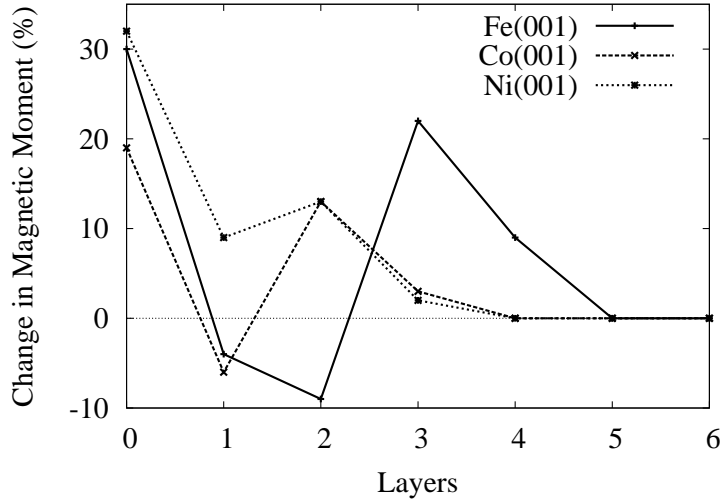}
\caption{Percentage variation of magnetic moment with respect to the bulk value for different layers.}\label{fig9}
\end{figure}

Table \ref{tab4}, shows magnetic moment approaches to the bulk value at S-5$^{th}$ layer in case of Fe(001) and at S-4$^{th}$ layer in the other two cases as expected from DOS result. Magnetic moments for different layers exhibit Friedel oscillation. This is also shown in the figure \ref{fig9}, layer based percentage change in magnetic moment compared to the bulk value. Figure \ref{fig9} further shows that the enhancement of surface magnetic moment is same for Fe(001) and Ni(001) but it is less for Co(001). Table \ref{tab4} also shows that d-electrons contribute most to the magnetic moment compared to s and p electrons as expected form DOS. Figure \ref{fig6} shows that at the Fermi level, difference between surface spin-up DOS and that of bulk is very small for Fe(001) and Co(001) and negligible for Ni(001). But there is significant difference between spin-down surface DOS and that of bulk DOS. This shows spin-down states are mainly responsible for the enhancement of magnetic moment at the surface. In all the three cases, the contribution of s- and p-orbital is comparable to TB-LMTO Green's function method \cite{alden}.
 
For comparison we have summarized our results of magnetic moments along with other available calculated and experimental values in table \ref{tab5}. Table \ref{tab5} shows that the enhancement of surface magnetic moment compared to the bulk is 30\% for Fe(001), 19\% for Co(001) and 32\% for Ni(001). Our result of bulk magnetic moment for Fe is within 5\% difference from experimental value \cite{danan, pauling}. Since enhancement of surface magnetic moment is less for Co(001) than the other two, therefore it is more stable to the change in the environment.

\begin{table}
\caption{Comparision of magnetic moment in $\mu_B$/atom for surface (S), sub-surfaces (S-1, S-2, S-3) and central layer or bulk (C/B). Number in the square brackets represents the reference numbers.}\label{tab5}
\footnotesize
\centering
\begin{tabular}{||c||c|c|c|c||c|c|c|c|c||c|c|c|c|c||}
\hline
& \multicolumn{4}{c||}{Fe(001)}  & \multicolumn{5}{c||}{Co(001)} &\multicolumn{5}{c||}{Ni(001)} \\ \cline{2-15}
\multicolumn{1}{||c||}{\raisebox{1.2ex}[0pt]{Methods}} & S & S-1 & S-2 & C/B & S & S-1 & S-2 & S-3 & S-4/B & S & S-1 & S-2 & S-3 & S-4/B\\ \hline

 & 2.98 & 2.35& 2.39 & 2.20 &1.86  & 1.64 & 1.65 & 1.64 & 1.65 & 0.68 &&&0.56 & \\ 
       & \cite{ohnishi,freeman,wu}&\cite{ohnishi}&\cite{ohnishi}&\cite{ohnishi}&\cite{li}&\cite{li}&\cite{li} & \cite{li}&\cite{li}&\cite{freeman} &&&
\cite{freeman} & \\ \cline{2-15}

FP-  &2.80 & 2.38 & 2.43 &  2.15&&&&&&0.73&0.68& 0.66& 0.63&\\
LAPW  &\cite{freeman1} & \cite{wu} & \cite{wu} & \cite{freeman}&&&&&&\cite{wu} &\cite{wu} & \cite{wu}& \cite{wu}&\\\cline{2-15}

  &&&& 2.30  &&&&&& 0.68 & 0.60 & 0.59 & 0.56 &\\
  &&&& \cite{wu} &&&&&& \cite{wimmer} & \cite{wimmer}& \cite{wimmer} &  \cite{wimmer}&\\ \hline

LMTO  & 2.87 & 2.34 & 2.33 & 2.18 &&&&&& 0.59 & 0.58 & 0.57 & 0.55 &\\
&\cite{eriksson,eriksson1}&\cite{eriksson,eriksson1}&\cite{eriksson,eriksson1}&\cite{eriksson,eriksson1}&&&&&&\cite{eriksson,eriksson1}&\cite{eriksson,eriksson1}&\cite{eriksson,eriksson1}&\cite{eriksson,eriksson1}& \\\hline

 & 2.97 & 2.30 & 2.37 & 2.24 & 1.84 & 1.63 & 1.66 & 1.64 & &0.69 & 0.64 & 0.66 & 0.64 & \\ 
 & \cite{alden} & \cite{alden} & \cite{alden} & \cite{alden} &  \cite{alden}& \cite{alden} &  \cite{alden}& \cite{alden} & &\cite{alden} &  \cite{alden}&\cite{alden} &  \cite{alden}& \\ \cline{2-15}

 & 2.97 & 2.30  & 2.37  & 2.25 & 1.84& 1.63&1.66&1.65&1.66&0.69& 0.64&0.66&0.64&0.65\\
 & \cite{niklasson}& \cite{niklasson} & \cite{niklasson} & \cite{niklasson} & \cite{niklasson}& \cite{niklasson}&\cite{niklasson}&\cite{niklasson}&\cite{niklasson}&\cite{niklasson}& \cite{niklasson}&\cite{niklasson}&\cite{niklasson}&\cite{niklasson}\\\cline{2-15}

TB-  & 2.86& 2.16& 2.38& 2.17& 1.76 & 1.46 & 1.58 & 1.56& 1.58  & 0.65 & 0.53 & 0.61 & 0.60 & 0.59 \\
LMTO  & \cite{biplab}& \cite{biplab}& \cite{biplab}& \cite{biplab}& \cite{monodeep} & \cite{monodeep} & \cite{monodeep} & \cite{monodeep}& \cite{monodeep}  & \cite{monodeep}& \cite{monodeep} & \cite{monodeep} & \cite{monodeep} & \cite{monodeep}\\\cline{2-15}

  & 2.98      & 2.17        &  2.40      & 2.26      &&&&&&&&&&\\ 
 &\cite{huda1}& \cite{huda1} & \cite{huda1}&\cite{huda1}&&&&&&&&&&  \\ \cline{2-15}

 & 2.95 &2.20 & 2.39  & 2.28  & &&& &&&&&&\\
 & \cite{monodeep} &\cite{monodeep} & \cite{monodeep}  & \cite{monodeep}  & &&& &&&&&&\\\cline{2-15}

 & 2.99 &2.21 & 2.38  & 2.26  & &&& &&&&&&\\
 & \cite{huda1} &\cite{huda1} & \cite{huda1}  & \cite{huda1}  & &&& &&&&&&\\ \hline

LCAO & 3.01& 1.69& 2.13& 1.84&&&&&& 0.44 & 0.58 & 0.62 & 0.56&0.54 \\
 & \cite{wang}& \cite{wang}& \cite{wang}& \cite{wang}&&&&&& \cite{wang1} & \cite{wang1} & \cite{wang1} & \cite{wang1}&\cite{wang1} \\\hline

{\bf ASR} & 2.99 & 2.17 & 2.38& 2.27 &&&&&&&&&& \\ 
 & \cite{biplab} & \cite{biplab} & \cite{biplab}& \cite{biplab} &&&&&&&&&& \\ \hline
 
 & & & & 2.21&&&&& 1.71&&&&& 0.616 \\
{\bf EXPT} & & & & \cite{danan}&&&&&\cite{pauling} &&&&&\cite{danan}  \\\cline{2-15}

 &&&& 2.22&&&&&&&&&&\\
 &&&& \cite{pauling}&&&&&&&&&&\\\hline

{\bf OUR} & 2.83 & 2.08 & 1.97 & 2.17& 1.83 & 1.45 & 1.74 & 1.59 & 1.54& 0.68 & 0.58 & 0.60 & 0.54 & 0.52\\ 
{\bf WORK} & &  &  & &  &  &  &  & &  &  &  &  & \\ \hline

\end{tabular} 
Note that Refs \cite{wimmer,alden,niklasson} use Green's function techniques; Ref \cite{biplab} is with TB-LMTO including surface dilatation; Refs \cite{huda1,biplab} are Supercell calculation; Refs \cite{huda1,monodeep} use Real space recursion.

\end{table}

Experimental values of local magnetic moment, like surface layer and layers below are not available for comparison. It is only the bulk magnetic moment for which experimental result is available. Work function is a surface property and is related to the surface DOS. It is an experimentally measurable quantity and therefore theoretical calculation of it is necessary for comparison with experimental result to test the accuracy of theoretical study. Table \ref{tab6} shows our calculated work functions for all the systems under consideration agree quite well with experimental and other theoretical calculations.\\

\begin{table}
\caption{Work functions of Fe(001), Co(001) and Ni(001) in eV. Number in the square brackets represents the reference numbers. Note that Ref \cite{alden} is with Green's function; Refs \cite{freeman2,ohwaki,ingle} use surface embedded Green's function.}\label{tab6}
\centering
\begin{tabular}{||c||c||c||c||}
\hline
Methods & Fe (001) & Co(001) & Ni(001) \\ \hline
 &4.29 \cite{ohnishi} & 5.17 \cite{li} & 5.37 \cite{wimmer},\\
FPLAPW&&& 5.5 \cite{freeman2},\\
&&&5.31 \cite{ohwaki}  \\ \hline
LAPW & & & 5.71 \cite{ingle}\\\hline
LMTO &4.30 \cite{eriksson1} &  & 5.02 \cite{eriksson1}\\ \hline
TB-LMTO & 4.5 \cite{alden} & 5.52 \cite{alden} & 5.75 \cite{alden}\\\hline
Experimental &4.67 \cite{michael}, &5.0 \cite{michael}  & 5.22 \cite{michael}\\ 
&4.4 \cite{tuner} &&\\\hline
{\bf Our work} &4.15 &5.33  &4.79 \\ \hline
\end{tabular} 
\end{table}

\section{Conclusion}

The augmented space formalism coupled with the recursion method and TB-LMTO has been successfully applied to study the effects of roughness on surface properties of bcc Fe(001), fcc Co(001) and fcc Ni(001). Two types of roughening are considered here, one with 10\% \& 20\% roughening only at the top layer and the other a more realistic surface, i.e., by roughening the first four layers with 20\%, 15\%, 10\%, \& 5\% from top layer respectively. Twelve atomic layers are considered to fully achieve bulk properties. Two layers of empty spheres are considered above the surface to include the vacuum. It is found that the trend in the variation of the width of DOS among the layers changes when a realistic surface is considered in comparison to a smooth surface. The comparison of surface properties for two different types of roughed surfaces shows slight difference in surface magnetic moment of Fe(001), whereas significant difference is obtained for Co(001) and no change is in the case of Ni(001) between these two types of surfaces. The  magnetic moment at the top layer is maximum in the cases of Fe(001) and Co(001) but it is at the S-3$^{rd}$ layer in the case of Ni(001) in both types of roughening. Layered based magnetic moment is also found different in both types of surfaces. Work functions of the respective systems found to be almost same for both types of rough surfaces as well as for the smooth surface. In all the cases, d-band has significant contribution towards the magnetic moment and there is significant surface effects on the d-band DOS in all the systems. We have shown that Augmented space formalism which works well for rough surfaces can also be applied to the nearly smooth surface. Therefore, we have carried out detailed study of the same systems for smooth surfaces also. Our result agree quite well with other available results. Layer wise magnetic moments show Friedel oscillations. The work function is found to be 4.08 eV, 5.30 eV and 4.76 eV for Fe(001), Co(001) and Ni(001) respectively. These values are closed to the available experimental and other calculated results.

\section*{Acknowledgments}
Authors would like to acknowledge Defense Research and Development Organization, India for providing financial support to this work under grant number ERIP/ER/1006016/ M/01/1373. P Parida acknowledges INSPIRE program division, Department of Science and Technology, India for providing a Senior Research Fellowship. We would like to thank Prof. O.K. Anderson, Max Plank Institute, Stuttgart, Germany, for his kind permission to use TB-LMTO code developed by his group.

\bibliographystyle{unsrt}
\bibliography{final}

\end{document}